\documentclass[aps, pra, twocolumn, superscriptaddress, showpacs]{revtex4}
\usepackage{amsmath,amssymb,graphicx,verbatim}
\usepackage[dvipdfm,colorlinks=true, citecolor=blue, urlcolor=blue, linkcolor=blue ]{hyperref}
\renewcommand{\section}[1]{{\par\it #1.---}\ignorespaces}
\begin{document}

\title{Simulating Zeno physics by quantum quench with superconducting circuits}
\author{Qing-Jun Tong}\email{tongqingjun0902@gmail.com}
\affiliation{Center for Interdisciplinary Studies $\&$ Key Laboratory for Magnetism and Magnetic Materials of the MoE, Lanzhou University,
Lanzhou 730000, China}
\affiliation{Center for Quantum Technologies, National University of Singapore, 3 Science Drive 2, Singapore 117543, Singapore}
\author{Jun-Hong An}\email{anjhong@lzu.edu.cn}
\affiliation{Center for Interdisciplinary Studies $\&$ Key Laboratory for Magnetism and Magnetic Materials of the MoE, Lanzhou University,
Lanzhou 730000, China}
\affiliation{Center for Quantum Technologies, National University of Singapore, 3 Science Drive 2, Singapore 117543, Singapore}
\author{L. C. Kwek}
\affiliation{Center for Quantum Technologies, National University of Singapore, 3 Science Drive 2, Singapore 117543,
Singapore}
\author{Hong-Gang Luo}
\affiliation{Center for Interdisciplinary Studies $\&$ Key Laboratory for Magnetism and Magnetic Materials of the MoE, Lanzhou University,
Lanzhou 730000, China}
\affiliation{Beijing Computational Science Research Center, Beijing 100084, China}
\author{C. H. Oh}\email{phyohch@nus.edu.sg}
\affiliation{Center for Quantum Technologies, National University of Singapore, 3 Science Drive 2, Singapore 117543,
Singapore}

\begin{abstract}
Studying out-of-equilibrium physics in quantum systems under quantum quench is of vast experimental and theoretical interests. Using periodic quantum quenches, we present an experimentally accessible scheme to simulate the quantum Zeno and anti-Zeno effects in an $\textit{open}$ quantum system of a single superconducting qubit interacting with an array of transmission line resonators. The scheme is based on the following two observations: Firstly, compared with conventional systems, the short-time non-exponential decay in our superconducting circuit system is readily observed; and secondly, a quench-off process mimics an ideal projective measurement when its time duration is sufficiently long. Our results show the active role of quantum quench in quantum simulation and control.
\end{abstract}
\pacs{03.67.Ac., 03.65.Xp, 03.67.Pp, 85.25.-j}
\maketitle

\section{Introduction}
Recently, there has been an increasing interest in exploring quench dynamics in quantum systems, in which a parameter of the system is tuned abruptly at specific time \cite{Polkovnikov2011}. Interest in this area has been evoked principally by several experimental realizations such as in ultra-cold atoms \cite{coldatom1,coldatom2}, where the high controllability in tuning parameters of the system has been achieved and the realization of different external sudden interruptions has been made accessible. Many important issues, for example the equilibration process and the relationship between thermalization and the integrability of many-body systems under quantum quench \cite{Rigol2008,Calabrese2006,Cazalilla2006,Sotiriadis2009,Calabrese2011,Kollath2007,Caux2012,Dorner2012,Daley2012}, have been investigated. So far, most of these works have focused on the ideal closed quantum systems. However, any realistic quantum system would become open due to the inevitable interaction with its surrounding environment, causing the so-called decoherence \cite{Breuer2002}. Studying quench dynamics in {\it open} system can not only cast better understanding on environment-induced decoherence effects under external interruptions but also give help in exploring potential strategies to control decoherence \cite{Nielsen2000}.

Decoherence control is one of the crucial issues in quantum information science \cite{Nielsen2000,Ladd2010}. Indeed, recent years have witnessed rapid progresses in decoherence-control strategies including quantum error correction codes \cite{Shor1995,Cory1998,Chiaverini2004,Schindler2011,Reed2012}, decoherence-free subspaces \cite{Duan1997,Lidar1998,Kwiat2000,Viola2001,Monz2009,Xu2012} and dynamical decoupling \cite{Viola1999,Uhrig2007,Du2009,Lange2010,Bylander2011,Piltz2013}. Quantum Zeno effect (QZE) and anti-Zeno effect (AZE) predicted by quantum mechanics are also quantum control strategies \cite{Facchi2008}. QZE states that the decay of an unstable quantum system can be slowed down by frequent measurements \cite{Misra1977}. The essential physics behind is that the frequent measurements make the decay of the quantum system follow periodically its short-time non-exponential (quadratic) behavior, preventing it from evolving into the exponential stage \cite{Facchi2008}. Moreover, it was predicted that an enhancement of decay due to frequent measurements could also be observed, a phenomenon which is called  AZE \cite{Kofman2000,Facchi2001}. Although shown theoretically to be possible for a variety of potential applications \cite{Erez2008,Maniscalco2008,Prezhdo2000,Huang2012,Paz-Silva2012,Kilina2013,Fujii2010}, QZE and AZE are not seen as prevalent a subject as the other methods that were mentioned above. This is mainly because that their experimental realizations have been beset with the following two obstacles: Firstly, the time scale for the initial non-exponential decay is often too short to implement frequent measurements, for example, this time scale for the spontaneous emission of atoms in the vacuum is roughly of the order of  $10^{-17}$s. Secondly, it is hard to realize frequent measurements on a highly unstable quantum system \cite{Facchi2008}.

In this Letter, using periodic quantum quenches, we propose an experimentally feasible implementation of QZE and AZE in a superconducting qubit (SQ) system interacting with a coupled transmission line resonator (TLR) array, as schematically shown in Fig.~\ref{fig1}(a). Our work is motivated by recent experimental achievements in the circuit QED system \cite{Clarke2008,You2011,Houck2011}, such as the tunable coupling between the SQ and the TLR \cite{Sillanp2007,Hofheinz2008,Neeley2010,Allman2010} and large scaled TLR lattices \cite{Houck2012,Houck2011}.  The high tunability in circuit QED systems has made it an ideal platform to study the quench dynamics in open quantum system. In the following, we show that the time scale of the non-exponential decay in this system, being of the order of nanoseconds, is typically much longer than that of a natural atom in the free space. More importantly, we show that a well-designed quantum quench protocol can simulate the ideal projection measurement very well.

\section{The model and its decoherence dynamics}
\begin{figure}[tbp]
\centering
\includegraphics[scale=0.95]{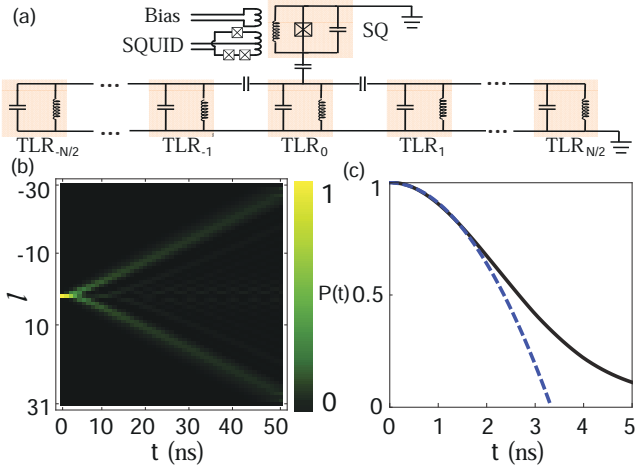}
\caption{(Color online) (a) Schematic illustration of the system consisting of an SQ interacting with a coupled TLR array. (b) Time evolution of the population distribution over the SQ-TLR system (SQ is located at the central TLR). (c) Detailed short-time non-exponential decay behavior (black solid line) of SQ in (b). The blue-dashed line is a numerical fit for $P(t)=1-(\frac{t}{\tau_z})^2$ with $\tau_z=3.33$ns. The parameters used are $\omega_0/2\pi=\omega_c/2\pi=8.74$GHz, $J/2\pi=g_0/2\pi=50$MHz and $N=60$.} \label{fig1}
\end{figure}

We consider a superconducting circuit system
consisting of a single SQ coupled capacitively to an $N$-TLR array. The TLRs are coupled each other in a tunable manner either via capacitances, as shown in Fig.~\ref{fig1}(a), or via Josephson junctions, such as dc- and rf-SQUIDs \cite{Peropadre2013}. The system Hamiltonian reads $\hat{H}(t)=\hat{H}_0+\hat{H}_I(t)$, with static part
\begin{equation}
\hat{H}_0=\omega_0\hat{\sigma}_+\hat{\sigma}_- +\sum_{l}[\omega_c
\hat{a}_l^\dag \hat{a}_l-J(\hat{a}_{l+1}^\dag \hat{a}_l+\text{H.c.})],
\end{equation}
where $\hat{\sigma}_\pm$ are the transition operators of the SQ with a transition frequency $\omega_0$, $\hat{a}_l^{\dag}$ ($\hat{a}_l$) is the creation (annihilation) operators of the $l$-th TLR with a resonant frequency $\omega_c$, and $J$ is the hopping rate between adjacent TLRs. The renormalization of cavity frequencies due to the coupling between cavities can be reduced via properly choosing the system parameters \cite{Peropadre2013}. The controllable interaction between the SQ and the central, i.e, the $0$-th, TLR is described by the Jaynes-Cummings model, $\hat{H}_\text{I}(t)=g(t)(\hat{\sigma}_+ \hat{a}_0+\hat{\sigma}_-
\hat{a}_0^{\dag})$, where $g(t)$ is the tunable coupling strength. Here, we do not consider the ultra-strong coupling regime, so that the rotating wave approximation in our model holds \cite{solano}.

This superconducting system represents a typical open quantum system. To see this, one can employ the Fourier
transformation $\hat{a}_l=\sum_k e^{i k l}\hat{a}_k/\sqrt{N}$. The Hamiltonian for the system can be written as
\begin{equation}
\hat{H}(t)=\omega_0 \hat{\sigma}_+\hat{\sigma}_- +\sum_k[\epsilon_k \hat{a}_k^\dag
\hat{a}_k+\frac{g(t)}{\sqrt{N}}(\hat{\sigma}_+ \hat{a}_k+\text{H.c.})],\label{hamil}
\end{equation}
where $\epsilon_k=\omega_c-2J \cos{k}$ is the nonlinear dispersion
relation describing an artificial environment with a finite bandwidth $4J$ \cite{Zhou2008}. The equation of motion of the reduced density matrix $\rho (t)$ of the SQ is governed by \cite{Breuer2002},
\begin{eqnarray}
\dot{\rho} (t) &=&-i\Omega (t)[\hat{\sigma}
_{+}\hat{\sigma}_{-},\rho (t)]+\Gamma (t)[2\hat{\sigma} _{-}\rho
(t)\hat{\sigma}
_{+}  \notag \\
&&-\hat{\sigma} _{+}\hat{\sigma} _{-}\rho (t)-\rho (t)\hat{\sigma} _{+}\hat{\sigma}
_{-}],
\end{eqnarray}
where $\Omega (t)=-\text{Im}[\frac{\dot{c}_{0}(t)}{c_{0}(t)}]$ and
$\Gamma (t)=-\text{Re}[\frac{\dot{c}_{0}(t)}{c_{0}(t)}]$ are the time-dependent shifted frequency and decay rate. The $c_0(t)$ is the amplitude of the SQ during time evolution and can be obtained via numerically solving the Schr\"{o}dinger equation since we are only interested in the single-excitation Hilbert space \cite{Tong2010}.

To reveal the distinct characters of our system during decoherence, we now study the non-Markovian decoherence dynamics of the SQ in the absence of quantum quench by setting $g(t)=g_0$ as a constant. We use $61$ TRLs, which is realizable under current technologies \cite{Houck2012,Houck2011}, and the parameter values practical in the typical circuit QED (see the caption in Fig.~\ref{fig1}) in the numerical simulation. For the resonant case shown in Fig. \ref{fig1}(b), one can see that the survived population $P(t)$ of the initially excited SQ decays completely to zero in tens of nanoseconds. It is important to note that, because of the strong non-Markovian effect in the structured environment, the quadratic decay behavior in our system is clearly seen. We find from Fig.~\ref{fig1}(c) that the time scale of the quadratic decay is roughly of the order of nanoseconds, much larger than the one in the atomic spontaneous emissions ($10^{-17}$s). This long time scale of the quadratic decay is a key ingredient for achieving Zeno effects in the system. Actually, one can even prolong this time scale by decreasing the hopping rate $J$. For example, when choosing $J=g_0=10$MHz, $\tau_z$ for the quadratic decay in Fig. 1(c) can reach $15$ns. However,  we cannot make this time scale too large either, since we are also limited by the practical intrinsic decay of the SQ. For the off-resonant case, population trapping can be achieved when $\omega_0$ lies near the dispersion band edge due to the forming of an atom-photon bound state, which may also lead to the radiation trapping for an initially un-populated atom in the case of few-photon transport \cite{Longo,Werra}.

\begin{figure}[tbp]
\centering
\includegraphics[scale=1]{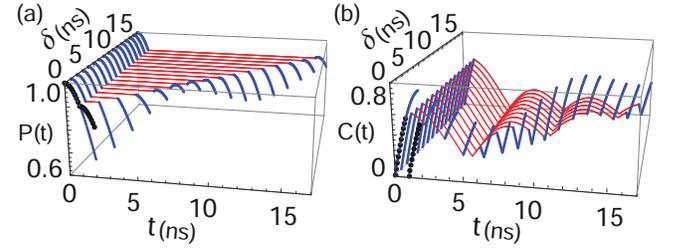}
\caption{(Color online) Time evolution of the survived population $P(t)$ of the SQ (a) and concurrence $C(t)$ between the SQ and its coupled TLR (b) under quench protocol (\ref{protocol}) for different quench-off time $\delta$. For comparison, we also plot the results under ideal projection measurement (black-thick dots). $\tau=1$ns and other parameters are the same as Fig.~\ref{fig1}(b).} \label{fig2}
\end{figure}
\section{Quench dynamics} With our controllable open quantum system, we can study its quench dynamics. In particular, we consider how the decoherence is affected when the coupling $g(t)$ is suddenly turned off for some time duration. Experimentally, the process of turning off the coupling can be efficiently achieved using a flux bias coil in phase qubit \cite{Sillanp2007,Hofheinz2008,Neeley2010,Allman2010}. We note that this decoupling has also been studied with transmon and flux qubits \cite{Gambetta2011,Srinivasan2011,Romero2012,Wang2009,Peropadre2010}. The following quench protocol is implemented:
\begin{eqnarray}
g(t) =\left\{ \begin{aligned} &g_0,\ \ \ \ \ \ \ \ \text{for}\ \ t \in
[0,\tau)\\ &0,\ \ \ \ \ \ \ \ \ \text{for}\ \ t\in [\tau,\tau+\delta)\\
&g_0,\ \ \ \ \ \ \ \ \text{for}\ \ t \in [\tau+\delta,\delta+2\tau]
\end{aligned}\right., \label{protocol}
\end{eqnarray}
where $\tau$ and $\delta$ are the quench-on and quench-off times, respectively.

Fig.~\ref{fig2}(a) depicts the quench dynamics in the dissipative SQ system.  As expected, the decay of the SQ is suddenly halted as soon as the coupling is quenched off. What happens when the coupling is quenched on again? We see that, depending on the quench-off time $\delta$, the dynamics during the second part of the time evolution process behaves differently. When $\delta=0$, i.e. the un-quench case, the decay in the second part is severer than the first one. This is because the decoherence dynamics is transferred from the short-time non-exponential to the subsequent exponential decay \cite{Fischer2009}. With the increasing of $\delta$, $P(t)$ in the second part shows a short-time increase, which means that the excitation transfers from the environment back to the SQ due to the strong non-Markovian backaction effect. Interestingly, with further increase of $\delta$, the decay of the SQ in the second part of the process behaves closer and closer to the first one, i.e., the system exhibits the quadratic decay.

Another interesting aspect we are concerned about is how the entanglement of the SQ and its coupled TLR behaves under the quantum quench. Using concurrence \cite{Concurrence}, we study  this entanglement under driving protocol (\ref{protocol}) in Fig.~\ref{fig2}(b). We find that initially the entanglement rapidly increases. If there is no quench, the entanglement goes on increasing. However, when the coupling is quenched off, the entanglement decreases substantially,  even though there are some oscillations when $\delta$ is too large. Actually, when the entanglement becomes sufficiently small, the state of the system approaches its initial state and the decay of the SQ would follow the quadratic behavior once the coupling is quenched on again, a picture which is consistent with the quench dynamics in Fig.~\ref{fig2}(a).

The above quench dynamics for large $\delta$ is reminiscent of the evolution under ideal projective measurement $\hat{\Pi}=\left\vert+\right\rangle\left\langle+\right\vert$ on the system, where the entanglement would decrease to zero suddenly due to the collapse of the wave function \cite{Breuer2002}. The dynamical evolution would subsequently trace exactly the same path as its initial stage does (see the black-thick dots in Fig.~\ref{fig2}). In this sense, the quench-off process for large $\delta$ in our scheme simulates a projective measurement on the dissipative SQ. Generally, the larger $\delta$ is, the better the quench-off process mimics a ideal projection measurement. However, one needs to be mindful that a large $\delta$ could also impose severe intrinsic decay of the SQ, so some compromises are necessary.

\section{Simulating Zeno physics}
In the above discussion, we have shown that the non-exponential decay in the system is observable under experimentally accessible conditions and the projective measurement can be simulated by quenching off the coupling for a sufficiently long time interval. With these ingredients, we expect that Zeno physics can be realized in this open SQ system by periodically quenching on and off the coupling to simulate the repetitive measurements. By choosing appropriate system parameters, the QZE and AZE can be identified according to whether the decay of the SQ during the total switch-on time is decelerated or accelerated compared with that of the un-quench case. Experimentally, one can design ultra-fast pulses via the flux bias coil to switch on and off the coupling periodically and read out the excited state population at the final stage using superconducting quantum interference device  \cite{Sillanp2007,Hofheinz2008,Neeley2010,Allman2010}.

\begin{figure}[tbp]
\centering
\includegraphics[scale=0.9]{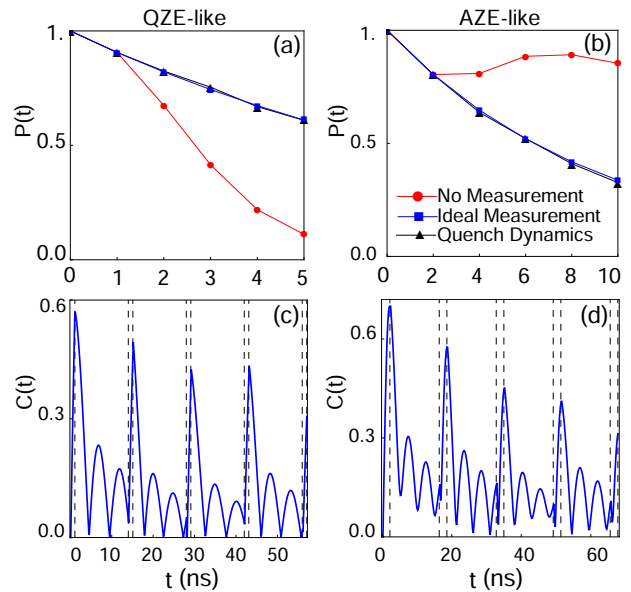}
\caption{(Color online)  (a) Simulating QZE: the survival population of the SQ for the case of no-measurement, ideal measurement, and periodic quantum quench when $\tau=1$ns, $\delta=13$ns, and $\omega_0/2\pi=8.74$GHz. (b): Simulating AZE when $\tau=2$ns, $\delta=14$ns, and $\omega_0/2\pi=8.54$GHz. Other parameters are the same as Fig.~\ref{fig1}(b). (c,d) are the entanglement dynamics in the whole quench process for (a,b) respectively.} \label{fig3}
\end{figure}
In order to simulate QZE, we consider the resonant case, in
which an excited SQ would decay to zero asymptotically in the absence of quantum quench.
In Fig.~\ref{fig3}(a), we plot the simulation result. Note that the quench-off processes are regarded as effective ``measurements" in our scheme and not shown in the plot. Compared with the no-measurement case (red-circled dots), in our scheme
(black-triangle dots) the decay process is indeed slowed down, which is
a clear signature of QZE. To further evaluate the efficiency of the method, we also calculate the decay process under the
ideal projection measurements (blue-squire dots). The survival population after $n$ ideal measurements is $P^{(n)}(t)=P(\tau)^n$ \cite{Facchi2008}.
One can see that our result matches very well with that of the later. The equivalence of these two methods has also been proved in a strict mathematical manner in Ref. \cite{Facchi2005}.

Fig.~\ref{fig3}(c) shows the dynamics of entanglement between the SQ and its coupled TLR in the whole evolution process. One can see that the entanglement decreases substantially in each time duration when the coupling is quenched off. As we had discussed above, because the SQ and its interacting TLR have become less entangled, the decay each time when the coupling is quenched on should follow its short-time non-exponential behavior. Hence, the total decay of the SQ is slowed down compared with the no-measurement case. Note that, in the conventional proposal of QZE \cite{Misra1977}, it is the measurement which disentangles the system and the environment and prevents the system from entering the exponential decay stage. Here this role is played by quantum quench.

To simulate the AZE, it is more convenient to consider the off-resonant case, in which the decoherence is partially suppressed due to the formation of a bound state in this dissipative SQ system \cite{Tong2010,John1990}. As shown in Fig.~\ref{fig3}(b), the population of the SQ in the absence of quantum quench is partially trapped and oscillates with time, manifesting strong non-Markovian effect. In the long-time limit, the decay rate would approach to zero \cite{Tong2010}. Because of this population trapping, the short-time decay of the SQ is severer than the long-time one. It is based on this observation that one can expect AZE if ideal projection measurements are implemented (see blue-squire dots). For our scheme, when we quench off the coupling several times, we also find that the decay of the SQ is accelerated. From Fig.~\ref{fig3}(d), we find that each time when the coupling is quenched off, the entanglement decreases, which implies that the SQ would follow its relatively severe decay behavior as in the initial stage when the coupling is switched on.
Interestingly, via simply tuning the decoherence process from partially
decoherence (quench-on) to non-decoherence (quench-off) periodically, we observe a completely
decoherence process. This interesting phenomenon can be well understood in terms of AZE.

\section{Discussion and conclusions} In terms of practical experiment, one should consider the intrinsic decay of both the SQ and TLRs. The intrinsic decay of SQ would lead to additional decoherence even in the quench-off processes. However, the whole dynamics in our scheme is implemented within $70$ ns, which means that, for a typical SQ with a life time of the order of microseconds \cite{Clarke2008,You2011,Houck2011}, this decoherence effect can be neglected.
About the decay of the TLRs, we find that it is beneficial for our scheme because it could make the SQ-TLR entanglement in the quench-off process approach zero more quickly.

Although we deal with a superconducting circuit system in this work, our results in studying quench dynamics in open quantum system are general. Actually, previous results obtained in both cold atomic \cite{Fischer2001} and optical systems \cite{Longhi2006} can be well understood in terms of the quench dynamics studied here. Other controllable quantum systems, like trapped ions \cite{Barreiro2011} and photonic crystals \cite{Noda2007}, can also be designed to study quench dynamics in open quantum system.

In summary, we have studied the quench dynamics in an open quantum system simulated by an SQ interacting with an array of coupled TLRs. The short-time non-exponential decay in this superconducting circuit system is prominent arising from the strong non-Markovian effect in the structured environment. We found that a long-time quenching-off process mimics very well an ideal projection measurement. Via quenching on and off periodically the coupling between the SQ and the TLR, we have simulated both QZE and AZE. The experimental feasibility of our scheme makes it a potential candidate to test the Zeno physics.  Our work sheds
some light on studying quench dynamics in open quantum system and exploring its novel applications in quantum control.

\section{Acknowledgement} We thank E. Solano and G. Romero for helpful discussions. The work is supported by the Fundamental Research Funds for the Central Universities, by the Specialized Research Fund for the Doctoral Program of Higher Education, by the NSF of China (Grants No. 11175072, No. 11174115 and No. 10934008), and by the National Research Foundation and Ministry of Education, Singapore (Grant No. WBS: R-710-000-008-271).

\end{document}